# When data sharing gets close to 100%: what ancient human DNA studies can teach the Open Science movement

Paolo Anagnostou[1,2], Marco Capocasa[2,3], Nicola Milia[4], Emanuele Sanna[4], Daniela Luzi[5] and Giovanni Destro Bisol[1,2*]

[1] Università di Roma ''La Sapienza'', Dipartimento di Biologia Ambientale, Rome, Italy

[2] Istituto Italiano di Antropologia, Rome, Italy

[3] Università di Roma ''La Sapienza'', Dipartimento Biologia e Biotecnologie "Charles Darwin", Rome, Italy

[4] Università di Cagliari, Dipartimento di Biologia Sperimentale, Cagliari, Italy

[5] Consiglio Nazionale delle Ricerche, Istituto di Ricerche sulla Popolazione e le Politiche Sociali, Rome, Italy

**\*Corresponding author:** Giovanni Destro Bisol

Dipartimento di Biologia Ambientale (sede di Antropologia), La Sapienza Università di Roma, Piazzale Aldo Moro 5, 00185 Rome, Italy

Fax: 0039649912771

e-mail address: destrobisol@uniroma1.it

**Abstract** – This study analyzes rates and ways of data sharing regarding mitochondrial, Y chromosomal and autosomal polymorphisms in a total of 162 papers on human ancient DNA published between 1988 and 2013. For the most part, data are available in such a way as to make their scrutiny and reuse possible. The estimated sharing rate is not far from totality (97.6% ± 2.1%) and substantially higher than observed in other fields of genetic research (Evolutionary, Medical and Forensic Genetics). A *questionnaire*-based survey suggests that the authors' awareness of the importance of openness and transparency for scientific progress is a fundamental factor for the achievement of such a high sharing rate. Most data were made available through body text, but the use of primary databases increased with the application of complete mitochondrial and next generation sequencing methods. Our study highlights three important aspects. First, we provide evidence that researchers' motivations are as necessary as stakeholders' policies and norms to achieve very high sharing rates. Second, careful analyses of the ways in which data are made available are an important first step to maximize data findability, accessibility, useability and preservation. Third and finally, the case of human ancient DNA studies demonstrates how Open Science can foster scientific advancements, showing that openness and transparency can help build rigorous and reliable scientific practices even in the presence of complex experimental challenges.

*Now is the right time to find examples of best practice; to celebrate these and to see what can be learnt from them.*

Neylon C and Wu S, Open science: tools, approaches, and implications, 2009.

**Introduction**

Making research data openly accessible to the scientific community is one of the main priorities for the global research system. In fact, there is wide consensus that data sharing may help scientific progress allowing a better exploitation of data and an optimized use of resources in a climate of scientific openness and transparency [1], [2]. However, there are also considerable barriers to be overcome, such as the inherent time and economic costs, possible data misuse, ethical issues and conflicts of interest with patenting discoveries [3], [4]. Given this tension, the diffusion of robust and effective open data practices should be viewed as an ongoing process which needs to be sustained by a cooperative effort of researchers, stakeholder and governments [2], [5], [6], [7], [8]. Strategies pursued by most academic institutions and funding bodies are mainly based on the development of digital infrastructures [9], [10] and policies for data sharing [5], [11], [12], while scientific journals encourage researchers to open their data through *ad hoc* guidelines [13], [14]. All these top-down initiatives are certainly indispensable. However, they may be empowered by bottom-up approaches such as empirical studies of data sharing practices based on *questionnaire*-based surveys or analyses of data retrievability from scientific literature [11], [15], [16]. Such initiatives may support the Open Science movement by providing quantitative answers to questions which regard norms (to what extent are they effective?), motivations (why researchers choose to share or withhold?) and sharing ways (do they comply with best practices?). Another significant outcome of this kind of study could be the identification of "flagship research fields", scientific areas of inquiry in which data sharing has already become common practice [17]. Apart from their symbolic value, identifying such positive examples may have a double outcome: (i) identify conditions and practices which may help spread data sharing; (ii) help understand whether and how data openness may play a role in the development of specific research fields. Unfortunately, studies carried out to date have not only failed to identify such "flagship" research fields, but also highlighted that data sharing is far from being common practice in all the research fields investigated so far [11], [18], [19], [20], [21], [22], [23], [24], [25], [26], [27], [28].

In this study, we analyze the rate and ways of data sharing in publications regarding human ancient DNA studies, a research field of particular interest for empirical studies due to its high standards in terms of reliability and experimental reproducibility. Combining a detailed analysis of published papers with a *questionnaire*-based survey, we finally show that data sharing is common practice in ancient human DNA studies and that such behaviour may be explained by the authors' awareness of the importance of openness and transparency for scientific progress. Thereafter, we compare the results obtained with findings of a previous study conducted in three genetic research fields (evolutionary, forensic and medical genetics) taking into consideration not only data availability but also the ways in which data are shared. Finally, we argue how the human ancient DNA case study might contribute to the Open Science movement, focusing on the importance of motivations to share and the need of looking carefully at the ways in which data are made available. Finally, we discuss the importance of openness and transparency in building rigorous and reliable scientific practices in the presence of complex experimental challenges.

**Methods**

*Data collection and analysis*

Our study is based on the scrutiny of papers published between October 1988 and December 2013, which were retrieved from the PubMed database (http://www.ncbi.nlm.nih.gov/pubmed) using 14 combinations of relevant key words (see Supplementary Table S1). The following species were considered: *Homo sapiens*, *Homo* *neanderthalensis* and *Homo denisovensis*. After removing irrelevant studies (e.g. studies not pertinent to human populations, reviews or meta-analyses), we selected 162 papers containing 133 mitochondrial, 29 Y chromosomal and 46 autosomal datasets. All papers were analyzed using an already developed protocol [29]. Further information regarding the experimental procedures (tissues collected, number of laboratories involved, independent replicates of raw data performed) is also provided as Supplementary Table S1.

Each paper went through two independent procedures of data collection, each performed by an experienced researcher. When conclusions were discordant, consensus was reached with the help of a third researcher who had independently analyzed the papers.

Datasets were counted as shared if they were presented in a way that permits their reuse in individual or population analyses without any substantial limitation:

- for unilinearly transmitted polymorphisms: when full haplotypic information of all individual DNAs genotyped and/or sequenced was available; this means that, when more than one type of polymorphism was analyzed (e.g. Single Nucleotide polymorphisms, SNPs, and microsatellites) it must be possible to reconstruct compound haplotypes.
- for autosomal polymorphisms: when the genetic profile for all loci genotyped/sequenced was made available for each individual analysed.

We identified three ways of withholding datasets (i) complete data unavailable (applicable only for unilinear polymorphisms): both SNP and microsatellites (or SNP and sequencing) haplotypic data were available, but the information needed to reconstruct compound SNP/microsatellites (or SNP/sequencing) haplotypes was not given; (ii) only part of data available: data were available but only for a part of individuals or of polymorphisms studied; (iii) only statistics-derived data available.

Datasets found to be shared went through further classification using four categories (body text, supplementary material, online databases and upon request). Differently from Milia et al.

[29], when a dataset was shared in more than one way (e.g. Online databases and supplementary material) only the most "effective" one was counted. Taking into account criteria of accessibility and preservation, depositing data in primary databases was regarded as the best practice, followed by supplementary material, open online repositories, body text and upon request (Supplementary Table S2). When a dataset was composed of two different types of markers shared in different ways (e.g. for mtDNA HVR1 sequences and coding region SNPs shared in online databases and body text, respectively), a value of 0.5 was assigned to each of them.

*Questionnaire based survey*

In order to gain further insights into the sharing behavior among researchers working with ancient human DNA, we asked first, last and corresponding authors of the papers inspected to answer some questions. Firstly, we collected information regarding their experience with ancient and modern DNA analysis. Secondly, we asked them to answer the following question: "Focusing on your overall publication experience, what is the contribution of the following factors to your choice of sharing ancient human DNA data?". Respondents were given the possibility to rate the following statements with four marks ("not important at all", "not very important", "important" and "very important): (i) Compliance with policies of scientific Journals, funding bodies or other stakeholders; (ii) Expectation to receive a higher number of citations; (iii) Awareness of the importance of making my own study open to scientific inquiry and (iv) Awareness that data sharing should be common practice which all researchers ought to comply with to foster scientific progress. Finally, we asked researchers to answer the question "What is the contribution of the following factors to the higher rate of data sharing in DNA studies of ancient compared to extant humans?" giving marks to the following statements: (i) The need to comply with more stringent policies of funding bodies and/or journals; (ii) The greater need to make data and results open to scientific inquiry; (iii) Lack or lesser weight of ethical/privacy constraints.

The survey was carried out using a Google docs, and no personal information was asked for in order to assure respondents anonymity. Of the 134 researchers emailed, 33 (24.0%) provided valid responses to the questionnaire.

*Data availability*

The datasets used for the analyses is provided as Supplementary table S1 and S3. Information regarding withholding papers is not reported here but is available upon request.

**RESULTS**

We inspected a total of 207 datasets regarding mitochondrial, Y chromosomal and autosomal polymorphisms, reported in 162 papers (published from 1998 to 2013) which had been selected using a key word driven Pubmed search. Mitochondrial datasets account for 63.8% of the total, and encompass SNP, Control region sequences and coding region/complete genomes. Y chromosomal datasets (13.5% of the total) comprise SNP and microsatellite polymorphisms. Finally, autosomal datasets (22.7%) include SNP, microsatellite and sequencing data, the latter being produced by next generation sequencing technologies (see Supplementary Table S4 for more details). The datasets predominantly regarded *Homo sapiens* (172, 83.1 % of the total) compared to *Homo neanderthalensis* (32, 15.5 %) and *Homo denisovensis* (3, 1.4 %; see Supplementary Table S5 for further details).

The yearly distribution of published datasets shows that since 1988, mtDNA has been, and still is, the most frequently used genetic system (figure 1). The use of autosomal and Y-chromosomal loci started to increase from 2003 and 2006, respectively.

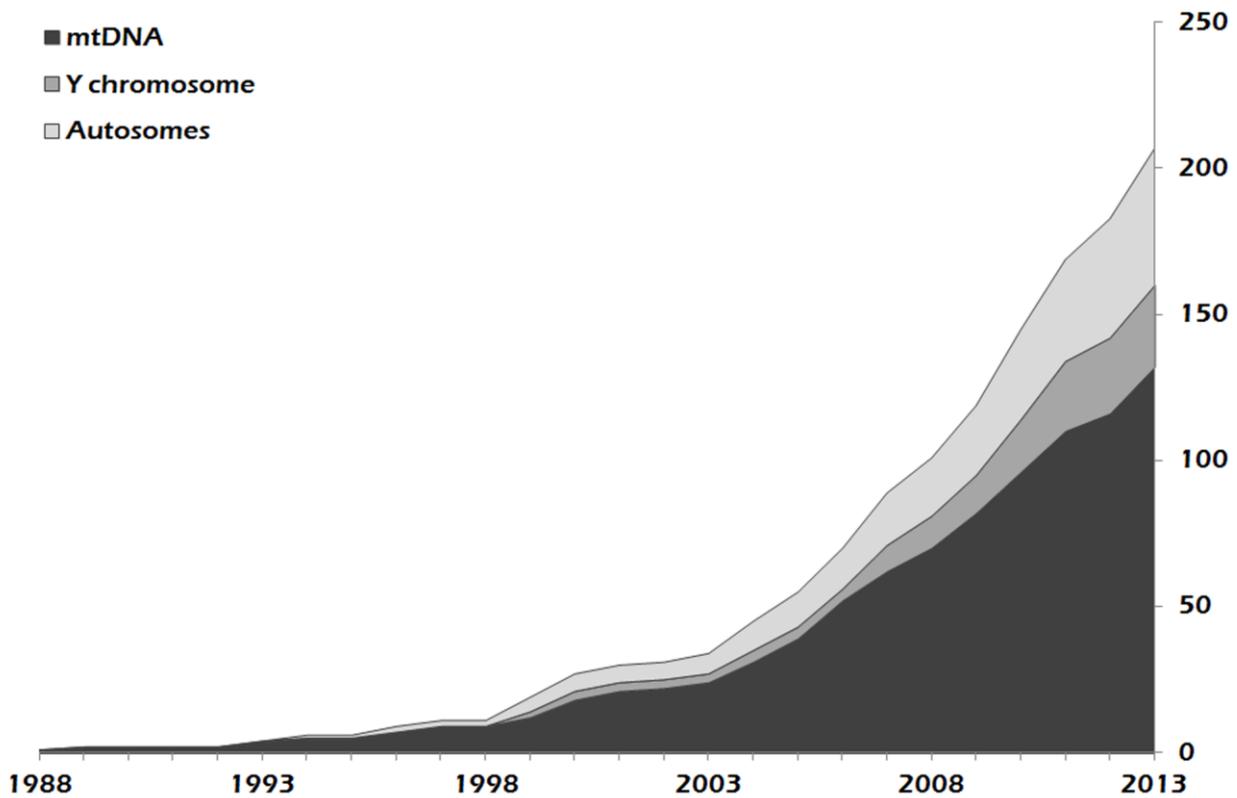

**Figure 1. Cumulative distributions of papers on human ancient DNA from 1988 since 2013 according to genetic system investigated.**

Two hundred and two datasets (97.6% ± 2.1%) were found to make their genetic information fully available and reusable (Table 1), with little variation among genetic systems (from 96.4% ± 6.9% for Y chromosome to 97.9% ± 4.1% for autosomes). No effective answer was received to emails sent to authors of withheld datasets, so that "immediate sharing" and after "email all authors" coincide. Presenting only data-derived statistics is the main way of withholding data. Interestingly, the five withheld datasets were published in the last six years: one dataset in 2008, two datasets in 2011 and two datasets in 2011. The resulting distribution does not comply with a decline of data availability with time since their publication, which fails to support the observation of Vines et al. regarding morphological data [30].

In addition to the estimates of sharing rates, we investigated how data are made available. It should be noted that we choose to consider all main ways of data sharing (body text, database, supplementary material, email requests), rather than focusing on a specific one (e.g. see [30], [31], [32]). In all genetic systems, datasets are more frequently shared using body text. In contrast with Y chromosome, a not negligible portion of mitochondrial and autosomal data is shared using online databases, mainly primary ones (e.g. Genbank) and, to a much lesser extent, open online repositories.

**Table 1. Data sharing ways in human ancient DNA studies.**

|  | mtDNA | Y chromosome | autosomes | total |
|---|---|---|---|---|
|  | (129:3)[a] | (27:1)[a] | (46:1)[a] | (202:5)[a] |
| *Shared datasets* | | | | |
| Online databases | 21,6% | - | 19.6% | 18.1% |
| Supplementary material | 21,6% | 29.6% | 27.1% | 23.8% |
| Open online repositories | - | - | 2.2% | 0.5% |
| Body text | 57.4% | 70,4% | 51,1% | 57.7% |
| *Withheld datasets* | | | | |
| Complete individual data unavailable | 33.3% | - | - | 20.0% |
| Only data derived statistics available | 66.7% | 100.0% | 100.0% | 80.0% |

[a] ratio between shared and withheld datasets

As a complement to the analysis of data retrievability from published papers, we asked the authors of inspected papers to give a mark concerning four possible factors that influence their decision on whether to share data or not. The vast majority of respondents indicated the importance of "making my own study open to scientific inquiry" (97.0% of respondents) and the awareness that "data sharing should be a common practice in scientific research" (93.9%) as the main reasons for making their data freely available to others. A slightly lower percentage (87.9%) pointed to the need to "comply with the sharing rule of Journals, funding bodies or other stakeholders" but only one third of them considers this as a very important factor regarding their choice to share. This is in line with the finding that a substantial part of papers (44.4%) was published in Journals in which data sharing is not mandatory. Finally, the expectation to receive a higher number of citations seems to have played only a minor role. On the whole, the sharing behaviour of respondents seems to be driven by epistemological motivations rather than external norms or expectations of any scientific reward.

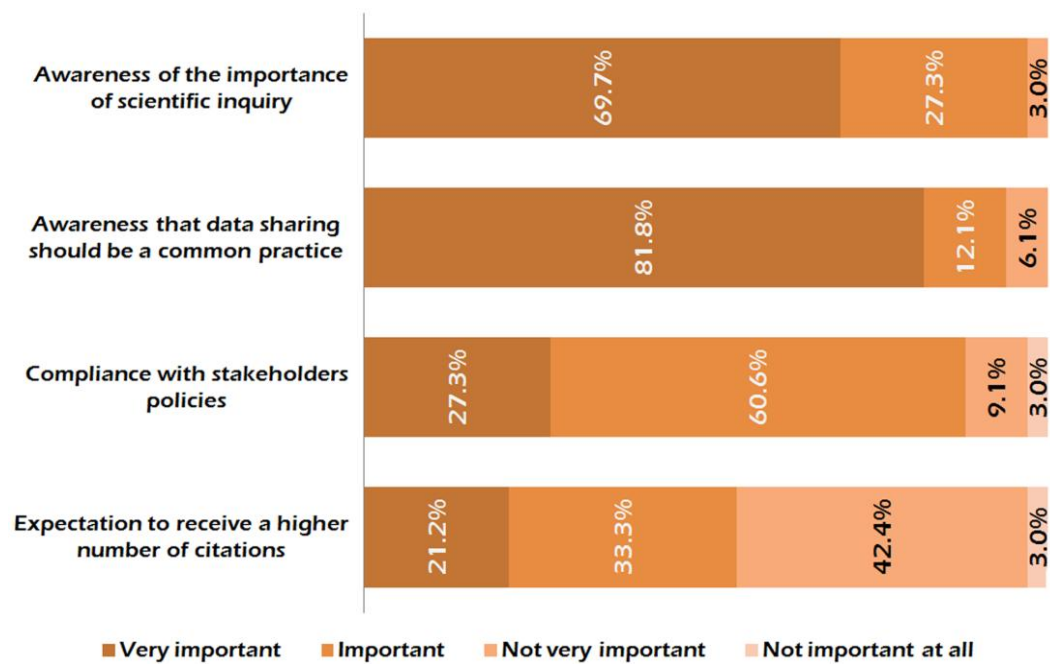

Figure 2. Frequency of responses to the question: "Focusing on your overall publication experience, what is the contribution of the following factors to your choice of sharing ancient human DNA data?". See Materials and Methods for complete statements.

## DISCUSSION

**Data sharing in different fields of genetic research**

In order to better appreciate the meaning of the results obtained in the course of this study, data for human ancient DNA studies were compared with those of Milia et al. [29] for evolutionary, forensic and medical Genetics. This comparison is particularly appropriate for two reasons. First, the two studies were carried out using the same criteria for paper selection, definition of "data", criteria to define shared and withheld datasets and following an identical workflow (see [29], pages 2-3). Second, the 4 research fields share not only most of their methodologies (based on DNA typing and sequencing), but also 3 important conditions which should favour data sharing: (i) the codified nature of genetic information; (ii) simplicity of basic metadata; (iii) availability of infrastructures for storage and dissemination. Thus, a number of confounding factors may be excluded.

As shown in figure 3, the sharing rate for human ancient DNA studies (recalculated to match exactly the genetic systems and period of data collection of Milia et al. [29]) is the highest and in two comparisons (with medical and evolutionary genetics) the difference is statistically significant (alpha = 0.05). Interestingly, all our values of sharing rates seem to be higher than estimates of data availability for other research fields (from 10%, [32] to 45% [15]). However, no true comparison may be carried out between our results and literature data since both the definition of data, inclusion criteria and workflow vary substantially among studies [15], [27], [28], [30], [31], [32].

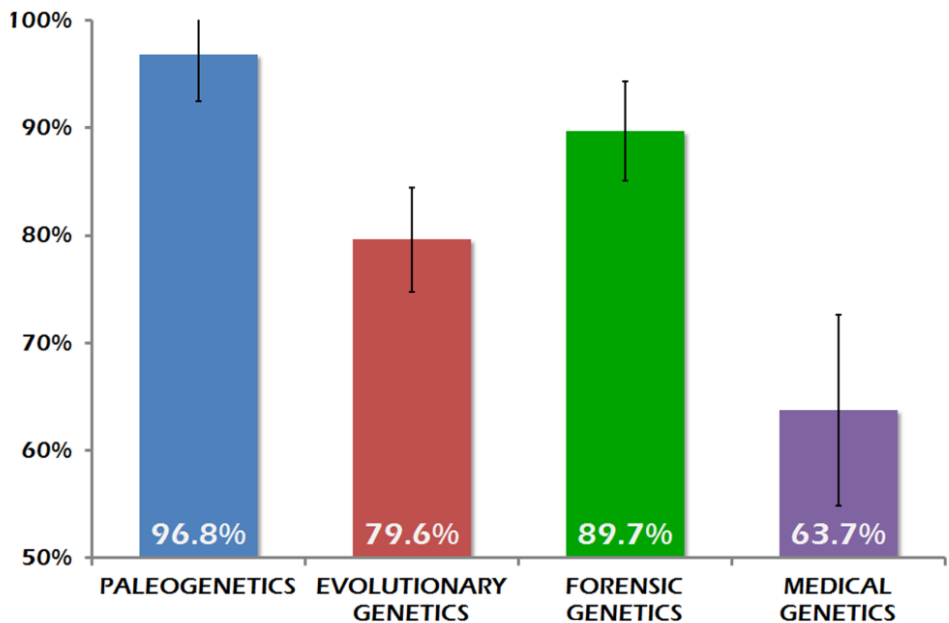

**Figure 3. Sharing rates in papers concerning mitochondrial and Y chromosomal polymorphisms (indexed in Medline from 1/1/2008 to 31/12/2011). Error bars indicate 95% confidence intervals.**

The results of the questionnaire-based survey turned out to be useful to gain insights into the difference observed in the sharing rate estimated in this study and in Milia et al. [29] (see figure 4). When we asked authors of surveyed papers that had also worked with extant populations (a total of 22 respondents) what reasons can explain the higher sharing rate of ancient DNA datasets, a large portion of respondents (84.8%) indicated "the greater need to make data and results open to scientific inquiry" as an important or very important factor. On the other hand, the answers "The need to comply with more stringent policies of funding bodies and/or journals" and "lack or lesser weight of ethical/privacy constraints", received lesser consideration, with 66.7% and 54.5% of respondents marking them as important or very important. Once more, the strong awareness of the importance of scientific inquiry seems be a key factor for scholars working on ancient human DNA.

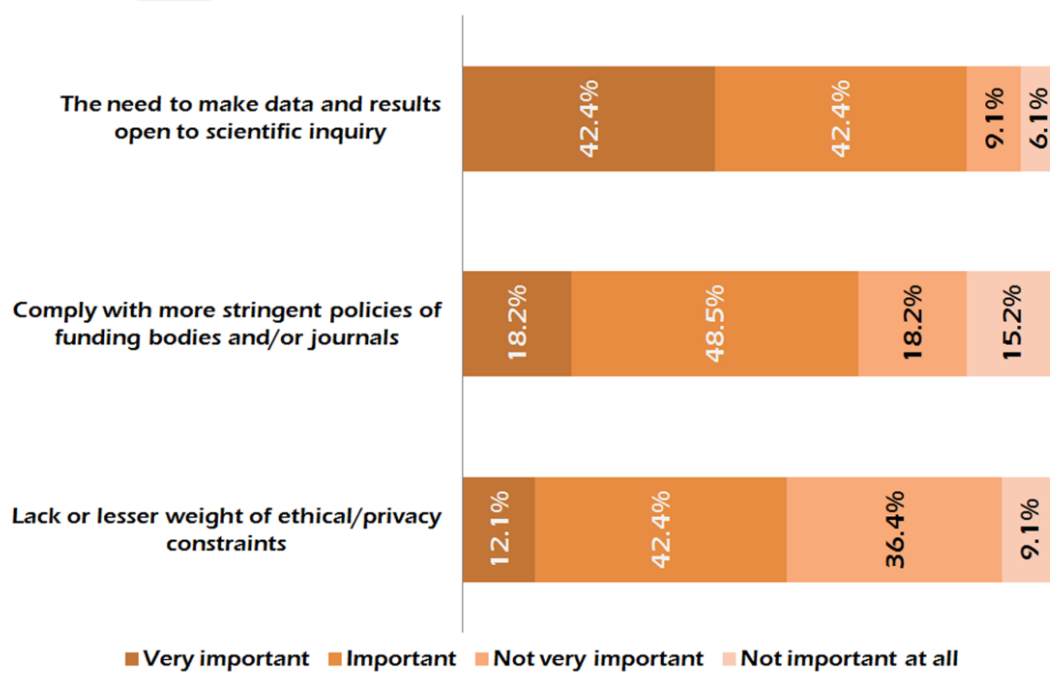

**Figure 4.** Frequency of responses to the question "What is the contribution of the following factors to the higher rate of data sharing in DNA studies of ancient compared to extant humans?". See Materials and Methods for complete statements.

Other useful insights are provided by the comparison of ways in which data are shared. As shown in figure 5, only in Medical Genetics we observed a more frequent use of body text (for both mtDNA Y and chromosome data) and a less frequent use of online databases than in human ancient DNA studies. Nonetheless, positive signals are observed when looking at the distribution of sharing ways from 1988 to 2013 (Supplementary Figure S1). In fact, it is evident that the use of online databases for mitochondrial and autosomal polymorphisms in human ancient DNA studies started to increase in 2006 and 2011, respectively - which coincides with the first application of complete mitochondrial and next generation sequencing in human ancient DNA studies - and their use prevails over other sharing ways in 2013. This trend is expected to continue in the future due to the likely increase in the use of new sequencing technologies, whose larger amount of data necessarily requires digital archiving.

However, to answer the question "how far we are from best practices" we should consider that in the studies taken into consideration, microsatellite and SNP polymorphisms (for all three genetic systems) were analyzed using methods which evaluate fragment length or allelic status at specific nucleotide positions, respectively. The resulting information cannot be deposited in primary databases since they are suitable for sequence data only. It follows that only a part of mtDNA can be deposited in primary databases (those obtained using sequencing methods), while using supplementary material becomes the best practice possible for Y chromosome data (which all refer to SNP and/or microsatellite polymorphisms). Therefore, to evaluate the situation more realistically, we calculated the ratio between the observed rate of compliance with the better practices and the maximum which could have been possible to achieve for each genetic system/research field (figure 6). The results obtained confirm the lack of congruence between rates and ways of sharing across research fields. In fact, the departure from best practice is substantial for both mtDNA and Y chromosome polymorphisms in human ancient DNA studies. On the other hand, evolutionary Genetics appears to be the field where the adopted data sharing ways ensure the highest degree of findability, accessibility, useability and preservation. An important implication of these results is that implementing the submission of microsatellite and SNP data in Genbank and interoperating databases is worth taking into consideration as a means to increase the compliance of data sharing ways with best practices, in particular for Y chromosome data.

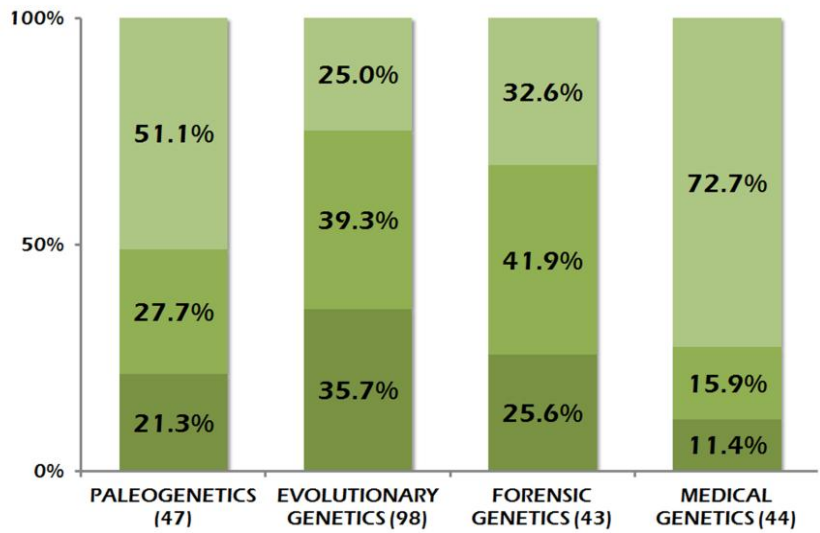
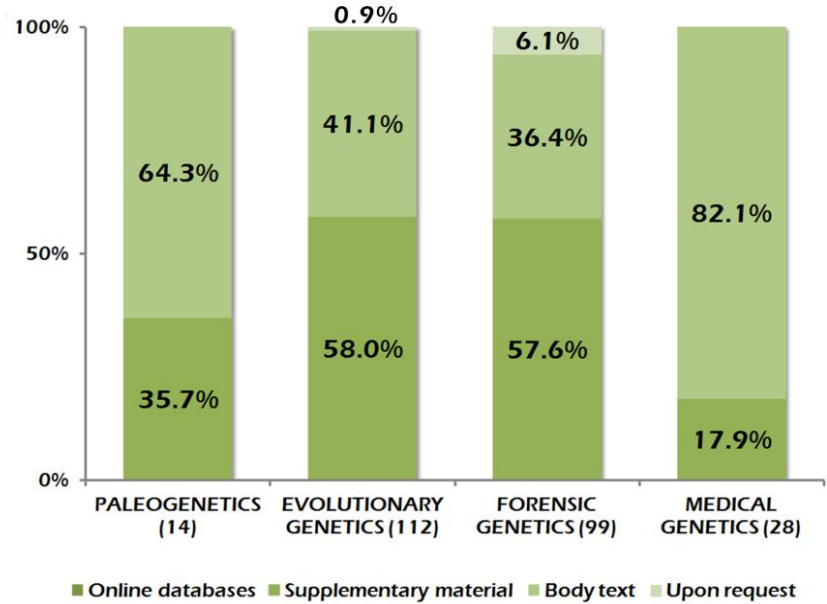

Figure 5. Frequencies of sharing ways in 4 genetic research fields based on the inspection of papers indexed in Medline from 01/01/2008 to 31/12/2011. The total number of scrutinized datasets for each field of research is reported in parentheses.

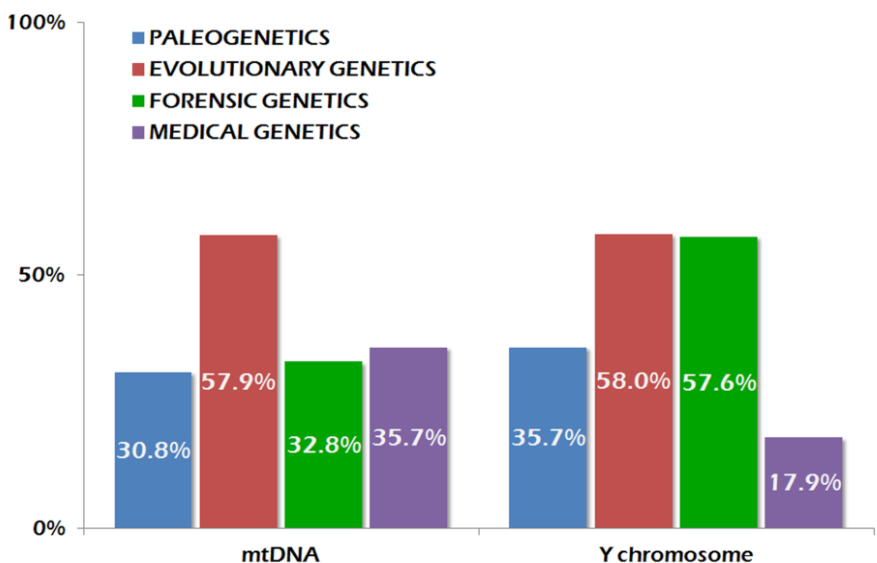

Figure 6. Ratio between the observed rate of compliance with the better practices and the maximum which could have been possible to achieve for mtDNA and Y chromosome, respectively.

**What ancient human DNA studies can teach the Open Science movement**

We believe that our analysis of data sharing in human ancient DNA studies conveys three important messages to all those who are interested in increasing the openness of research data.

First, we provided evidence that awareness of the importance of transparent scientific practices may be an indispensable complement to policies and rules of Journals, funding bodies and other stakeholders in order to achieve a very high data sharing rate. This points to the need to make all actors in scientific research conscious of the importance of open data to improve quality and reproducibility of research products [33]. We sustain that a key step is in the education of young researchers to the principles of open science, so as to make them understand its connections with scientific progress and appreciate the importance of transparency and trust in research [16], [34], [35], [36]. Human ancient DNA studies may serve as an excellent case study for these purposes.

Second, from what we observed for human ancient DNA studies, a very high sharing rate is not necessarily associated with the preferential use of archiving tools which make data more easily accessible, findable, useable and better preserved. Therefore, attention should be paid not only to the rate but also to the way in which data are shared. We have shown that by taking into account all the different ways of sharing data (body text, supplementary materials, open online repositories, open online databases and upon request), we may obtain a more complete assessment of the scientific practices and understand what the most important barriers are to a robust and effective data sharing. This latter point is well exemplified by the cause-effect relation between the unavailability of primary databases for specific types of genetic loci and the lack of compliance with best practices of data archiving in different fields of genetic research.

Third and finally, the case of human ancient human DNA studies provides an example of how data openness and transparency may play an important role in the development of specific research fields. This can probably be better understood by briefly looking at the history of human ancient DNA studies. Pioneered by Svante Pääbo [37] in mid 80's, this field immediately attracted great interest due to its potential in shedding light on key issues of human evolution [38]. However, its development was hampered by controversies surrounding the time of DNA preservation and the risk of contamination during excavations and laboratory procedures [39], [40]. In fact, the DNA sequences obtained from a 2,400-yr-old mummy by Pääbo [37] using molecular cloning is today considered to be a result of contamination [41]. More in general, the field of ancient human DNA was considered by many to be untrustworthy until the application of next-generation sequencing [42]. Nonetheless, human paleogenetics is today a small but absolutely vital research field, which takes advantage of next-generation sequencing techniques to increase its analytical power. This includes testing for contamination, and attracts particular interest from the scientific community and the public [43], [44]. We argue that openness of researchers to the scientific scrutiny of their data coupled with the adoption of stringent standards and cross-laboratory validation procedures has been crucial in overcoming doubts concerning scientific rigor and data reliability [41]. In this way, human ancient DNA studies avoided the decline which occurred with other promising approaches adopted to study the remote human evolutionary past, such as DNA-DNA hybridization [45], where lack of reproducibility was a critical aspect. Thus, the case of human ancient DNA studies illustrates that data sharing and, more in general, openness to scientific inquiry, can help build rigorous and reliable scientific practices even in the presence of complex experimental challenges.

## Acknowledgments

This work was supported by the Ministero dell'Istruzione, dell'Università e della Ricerca (PRIN 2009- 2011, prot.n. 200975T9EW) and the Istituto Italiano di Antropologia. We thank Eske Willerslev for help with an earlier draft of this paper.

## Author Contributions

G.D.B., P.A., M.C., N.M. and E.S. conceived the study. P.A., M.C. and N.M. collected and analyzed the data. D.L. helped with the *questionnaire* design and administration. G.D.B. wrote the paper. All authors read and approved the manuscript.

## Additional information

### Competing interests

The authors have no competing interests.

## References

1. Fischer J, Zigmond MJ (2010) The essential nature of sharing in science. Sci Eng Ethics 16: 783-79.

2. Boulton G, Campbell P, Collins B, Elias P, Hall W, et al. (2012) Science as an open enterprise. London: The Royal Society. 105 p.

3. Murdoch C, Caulfield T (2009) Commercialization, patenting and genomics: researcher perspectives. Genome Med 1: 22.

4. Giffels J (2010) Sharing data is a shared responsibility: Commentary on: «The essential nature of sharing in science». Sci Eng Ethics 16: 801-803.

5. National Institutes of Health (2003). NIH data sharing policy and implementation guidance. Available: http://grants.nih.gov/grants/policy/data_sharing/data_sharing_guidance.htm. Accessed 7 July 2014.

6. Biotechnology and Biological Sciences Research Council (2007). BBSRC data sharing policy: version 1.1 (June 2010 update). Available: http://www.bbsrc.ac.uk/datasharing. Accessed 7 July 2014.

7. Thorley M (2010) NERC data policy – guidance notes. Available: http://www.nerc.ac.uk/research/sites/data/documents/datapolicy-guidance.pdf. Accessed 7 July 2014.

8. Holdren JP (2013) Increasing access to the results of federally funded scientific research. Available: http://www.whitehouse.gov/sites/default/files/microsites/ostp/ostp_public_access_memo_2013.pdf. Accessed 7 July 2014.

9. Lecarpentier D, Michelini A, Wittenburg P (2013) The building of the EUDAT Cross-Disciplinary Data Infrastructure. EGU General Assembly Conference Abstracts 15: 7202.

10. Manghi P, Manola N, Horstmann W, Peters D (2010) An infrastructure for managing EC funded research output - The OpenAIRE Project. The Grey Journal 6: 1.

11. Tenopir C, Allard S, Douglass K, Aydinoglu AU, Wu L, et al. (2011) Data sharing by scientists: practices and perceptions. PLoS One 6: e21101.

12. University of Edinburgh (2014) Research data management policy. Available: http://www.ed.ac.uk/schools-departments/information-services/about/policies-and-regulations/research-data-policy. Accessed 7 July 2014.

13. Groves T (2010) BMJ policy on data sharing. BMJ 340: c564.

14. Whitlock MC, McPeek MA, Rausher MD, Rieseberg L, Moore AJ (2010) Data archiving. Am Nat 175: 145-146.

15. Piwowar H (2011) Who shares? Who doesn't? Factors associated with openly archiving raw research data. PLoS One 6: e18657.

16. Destro Bisol G, Anagnostou P, Capocasa M, Bencivelli S, Cerroni A, et al. (2014) Perspectives on Open Science and scientific data sharing: an interdisciplinary workshop. J Anthropol Sci 92: 179-200.

17. Neylon C, Wu S (2009) Open Science: tools, approaches, and implications. Pac Symp Biocomput 2009: 540-544.

18. Campbell EG, Clarridge BR, Gokhale M, Birenbaum L, Hilgartner S, et al. (2002) Data withholding in academic genetics: evidence from a national survey. JAMA 287: 473-480.

19. Blumenthal D, Campbell EG, Gokhale M, Yucel R, Clarridge B, et al. (2006) Data withholding in genetics and the other life sciences: prevalences and predictors. Acad Med 81: 137-145.

20. Vogeli C, Yucel R, Bendavid E, Jones LM, Anderson MS, et al. (2006) Data withholding and the next generation of scientists: results of a national survey. Acad Med 81: 128-136.

21. Alsheikh-Ali AA, Qureshi W, Al-Mallah MH, Ioannidis JPA (2011) Public availability of published research data in high-impact journals. PLoS One 6: e24357.

22. Wicherts JM, Bakker M, Molenaar D (2011) Willingness to share research data is related to the strength of the evidence and the quality of reporting of statistical results. PLoS One 6: e26828.

23. Enke N, Thessen A, Bach K, Bendix J, Seeger B, et al. (2012) The user's view on biodiversity data sharing – Investigating facts of acceptance and requirements to realize a sustainable use of research data. Ecological Informatics 11: 25-33.

24. Kim Y, Stanton JM (2012) Institutional and individual influences on scientists' data sharing practices. Journal of Computational Science Education 3: 47-56.

25. Hampton SE, Strasser CA, Tewksbury JJ, Gram WK, Budden AE, et al. (2013) Big data and the future of ecology. Frontiers in Ecology and the Environment 11: 156-162.

26. Luzi D, Ruggieri R, Biagioni S, Schiano E (2013) Data sharing in environmental sciences: a survey of CNR researchers. International Journal of Grey Literature

27. Caetano DS, Aisenberg A (2014) Forgotten treasures: the fate of data in animal behavior studies. PeerJ PrePrints 2: e396v1.

28. Magee AF, May MR, Moore BR (2014) The dawn of Open Access to phylogenetic data. arXiv 1405.6623.

29. Milia N, Congiu A, Anagnostou P, Montinaro F, Capocasa M, et al. (2012) Mine, yours, ours? Sharing data on human genetic variation. PLoS One 7: e37552.

30. Vines TH, Albert AY, Andrew RL, Débarre F, Bock DG, et al. (2014) The availability of research data declines rapidly with article age. Curr Biol 24: 94-97.

31. Wicherts JM, Borsboom D, Kats J, Molenaar D (2006) The poor availability of psychological research data for reanalysis. Am Psychol 61: 726–728.

32. Savage CJ, Vickers AJ (2009) Empirical study of data sharing by authors. PLoS One 4: e7078.

33. ALLEA (2012) Open Science for the 21st century. A declaration of ALL European Academies. Available: http://www.unesco.org/new/en/communication-and-information/resources/news-and-in-focus-articles/all-news/news/open_science_for_the_21st_century_declaration_of_all_european_academies/#.U7ppbb_O87B. Accessed 7 July 2014.

34. Barr CD, Onnela JP (2012) Establishing a culture of reproducibility and openness in medical research with an emphasis on the training years. Chance 25: 8-10.

35. Feldman L, Patel D, Ortmann L, Robinson K, Popovic T (2012) Educating for the future: another important benefit of data sharing. Lancet 379: 1877-1878.

36. Piwowar HA, Becich MJ, Bilofsky H, Crowley RS, on behalf of the caBIG Data Sharing and Intellectual Capital Workspace (2008) Towards a data sharing culture: recommendations for leadership from academic health centers. PLos Med 5: e183.

37. Pääbo S (1985) Molecular cloning of ancient Egyptian mummy DNA. Nature 314: 644-645.

38. Stoneking M, Krause J (2011) Learning about human population history from ancient and modern genomes. Nat Rev Genet 12: 603-614.

39. Hebsgaard MB, Phillips MJ, Willerslev E (2005) Geologically ancient DNA: Fact or artefact? Trends Microbiol 13: 212-220.

40. Pääbo S, Poinar H, Serre D, Jaenicke-Després V, Hebler J, et al. (2004) Genetic analyses from ancient DNA. Annu Rev Genet 38: 645-679.

41. Shapiro B, Hofreiter M (2010) Analysis of ancient human genomes: using next generation sequencing, 20-fold coverage of the genome of a 4,000-year-old human from Greenland has been obtained. Bioessays 32: 388-391.

42. Wall JD, Kim SK (2007) Inconsistencies in Neanderthal genomic DNA sequences. PLoS Genet 3: e175.

43. Kirsanow K, Burger J (2012) Ancient human DNA. Ann Anat 194: 121-132.

44. Fortes G, Speller CF, Hofreiter M, King TE (2013) Phenotypes from ancient DNA: Approaches, insights and prospects. Bioessays 35: 690-695.

45. Marks J, Schmid CW, Sarich VM (1988) DNA hybridization as a guide to phylogeny: relations of the Hominoidea. J Hum Evol 17: 769-786.

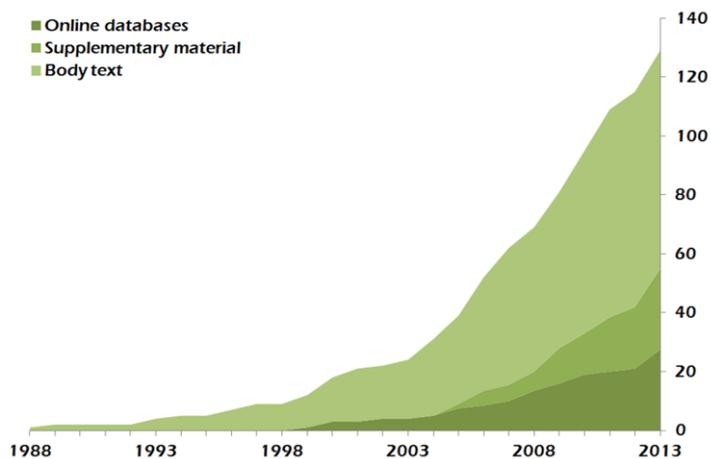
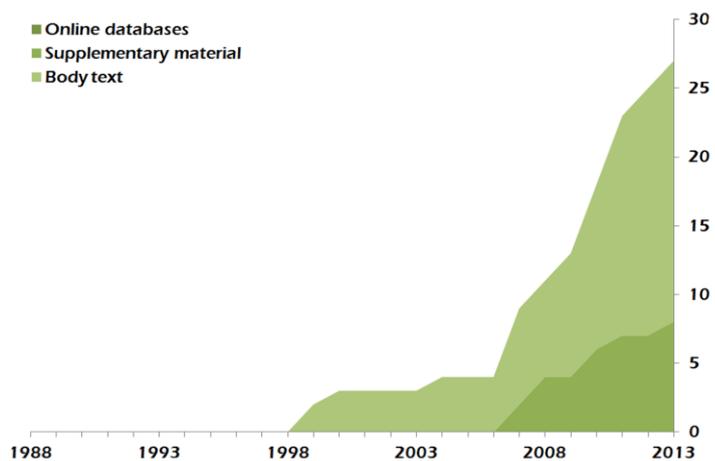
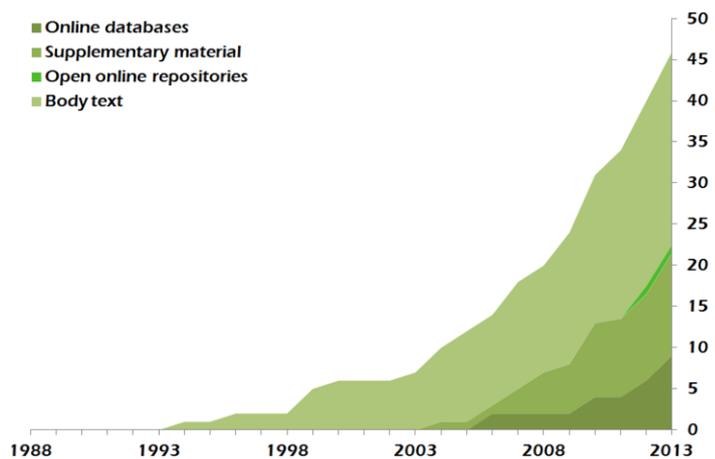

**Supplementary Figure S1.** Cumulative distributions of different sharing ways from 1988 since 2013 according to the genetic system investigated.

**Supplementary table S1.** Information collected on datasets analyzed in the course of this study. Na= information not available.

**Supplementary table S3.** Answers to the questionnaire.

**These tables are available upon request**

**Supplementary Table S2**. Efficacy of different ways of data sharing in terms of of findability, accessibility, useability and preservation.

| | Findability | | Accessibility | | Useability | | Preservation | |
|---|---|---|---|---|---|---|---|---|
| | Rate | notes | Rate | notes | Rate | notes | Rate | notes |
| Open online databases[1] | High | Standard indexing metadata formats allow an easy discovery of datasets regardless of the availability of the related paper | High | Free and immediate access to data (except if dataset is under embargo or access depend on the basis on informed consent) | High | Frequent use of standardized data and metadata formats | High | Systematic use of digital backup procedures (especially for primary databases which exchange data every day) |
| Supplementary material | Low | Conditioned to the discovery of the related paper | High | They are usually freely accessible even in pay per access papers. | Medium | Standardized data and metadata formats are not requested but often used. | Medium | Potential loss and/or unaivalability of datasets due to non working links to either publishers or authors web site (Anderson et al. 2006) |
| Open online repositories[2] | Low | Conditioned to the discovery of the related paper and to the presence of specific information therein. | High | Free and immediate access to data (except if access depend on the basis on informed consent) | Low/Medium | Depending upon the data and metadata format used by the author/s | Low/Medium | Data may be lost or unavailable due to link failure or data removal. |
| Body text | Low | Depending on to the discovery of the related paper | Low | Depending on the accessibility status of the related paper (open or pay per access) | Low | Need for data transformation before being able to use it (usually extensive and time consuming) | High | Digital and/or printed version of the paper should be always available. |
| Upon request | Low | Depending on the discovery of the related paper | Low | Conditioned to proper email working (Wren et al. 2006) and actual willingness to provide data. | Low/Medium | Depending upon the data and metadata format used by the author/s | Low | Data may be lost or unavailable due to computer failure or change of mail address. |

[1]. In this category all structured data archives that allow queries are included (e.g. Genebank)

[2]. In this category the archives that only store data not allowing any kind of query are included (e.g. institution sites, personal sites, data collections).

**References**

Anderson NR, Tarczy-Hornoch P, Bumgarner RE. 2006. On the persistence of supplementary resources in biomedical publications. BMC Bioinformatics. 7:260.

Wren JD, Grissom JE, Conway T. 2006. E-mail decay rates among corresponding authors in MEDLINE. The ability to communicate with and request materials from authors is being eroded by the expiration of e-mail addresses. EMBO Rep. 7:122-127.

**Supplementary table S4**. Characterization of the datasets under scrutiny in terms of genetic polymorphisms.

| | *Control region sequences (HVR1, HVR2 or both)* | *Coding region SNPs* | *Coding region/complete mithocondrial genome sequence* | *Control region sequences and coding region SNPs* | *Control region and Coding region/complete mithocondrial genome sequences* | *Total* |
|---|---|---|---|---|---|---|
| **mtDNA** | 45 | 7 | 11 | 66 | 3 | 132 |

| | *Microsatellites* | *SNPs* | *Microsatellites and SNPs* | *Total* |
|---|---|---|---|---|
| **Y chromosome** | 6 | 8 | 14 | 28 |

| | *Microsatellites* | *SNPs* | *Sequences* | *Microsatellites and SNPs* | *Total* |
|---|---|---|---|---|---|
| **Autosomal** | 23 | 11[a] | 10[b] | 3 | 47 |

[a] One dataset was genotyped with microarray technology.
[b] Mostly genotyped with Next generation sequencing technology (9 out of 10).

**Supplementary table S5.** Characterization of the datasets under scrutiny in terms of species of the *genus* Homo.

|  | mtDNA | Y chromosome | autosomes | Total |
|---|---|---|---|---|
| ***Homo sapiens*** | 110 | 26 | 36 | 172 |
| ***Homo neanderthalensis*** | 21[a] | 2 | 9 | 32 |
| ***Homo denisovensis*** | 1 | 0 | 2 | 3 |

[a] Includes the genome of the hominin from Sima de los Huesos.